\documentclass[12pt,aps,prb,preprint,superscriptaddress]{revtex4}

\usepackage[english]{babel}
\usepackage{url}

\def\reff#1{(\ref{#1})}
\newcommand{\be}{\begin{equation}}
\newcommand{\ee}{\end{equation}}

\def\smfrac#1#2{{\textstyle{#1\over #2}}}

\def\spose#1{\hbox to 0pt{#1\hss}}
\def\ltapprox{\mathrel{\spose{\lower 3pt\hbox{$\mathchar"218$}}
 \raise 2.0pt\hbox{$\mathchar"13C$}}}
\def\gtapprox{\mathrel{\spose{\lower 3pt\hbox{$\mathchar"218$}}
 \raise 2.0pt\hbox{$\mathchar"13E$}}}

\newcommand{\tfo}{ {{}_2 \! F_1} }

\begin{document}

\title{The falling raindrop, revisited}

\author{Alan D. Sokal}
\email{sokal@nyu.edu}
\affiliation{Department of Physics, New York University,
      4 Washington Place, New York, NY 10003, USA}
\affiliation{Department of Mathematics,
      University College London, London WC1E 6BT, United Kingdom}

\date{August 1, 2009}

\begin{abstract}
I reconsider the problem of a raindrop falling through mist, collecting mass,
and generalize it to allow an arbitrary power-law form for the accretion rate.
I show that the coupled differential equations can be solved
by the simple trick of temporarily eliminating time ($t$)
in favor of the raindrop's mass ($m$) as the independent variable.
\end{abstract}

\maketitle 


A perennial homework exercise in differential-equation-based courses
in Newtonian mechanics is the problem of a raindrop falling through mist,
collecting mass.\cite{Krane_81,Adawi_86,Dick_86,Partovi_89}
If the rate of accretion is assumed to depend only on
the raindrop's current mass (or radius) and not on its velocity,
then the solution is fairly straightforward.
When, by contrast, the rate of accretion is taken to depend
on both the mass and the velocity,
one is faced with a pair of coupled differential equations,
and the trick for disentangling them can be surprisingly difficult to find
--- not only for the student, but also for the instructor who has forgotten
the method after some years' absence and must rediscover it
(as I can personally testify from recent experience).\cite{note_quotes}

Here I would like to show that a very general version of
the raindrop problem [see \reff{eq.newton}/\reff{eq.accretion} below]
can be solved by using a versatile technique that ought to have a place in
all students' (and instructors') mathematical arsenals:
namely, eliminating reference to the old independent variable
(here the time $t$) and temporarily taking one of the
old dependent variables as the new independent variable.

Students may remember this trick from the analysis of
one-dimensional motion with a force that depends on position ($x$)
and velocity ($v$) but not explicitly on time:
\be
   m \, {dv \over dt}  \;=\;  F(x,v)   \;.
\ee
By using the chain rule
\be
   {dv \over dt}  \;=\;  {dv \over dx} \: {dx \over dt}  \;=\;
                         v \, {dv \over dx}
\ee
we can temporarily eliminate $t$ and instead take $x$ as the new
independent variable:  this yields the first-order differential equation
\be
   mv \, {dv \over dx}  \;=\;  F(x,v) 
 \label{eq.dvdx}
\ee
for the unknown function $v(x)$.
In some cases this equation can be solved explicitly.\cite{note_solve_Fxv}
Once one has in hand the function $v(x)$,
one can reinstate time and solve (at least in principle)
the first-order separable differential equation
\be
   {dx \over dt}  \;=\;  v(x)   \;.
\ee

Let us now consider the raindrop problem,
which involves a pair of coupled differential equations
for two unknown functions:
the raindrop's mass $m(t)$ and its velocity $v(t)$.
The first equation is the Newtonian equation of motion for the raindrop,
\be
   m \, {dv \over dt}  \:+\:  v \, {dm \over dt}   \;=\;  mg
   \;,
 \label{eq.newton}
\ee
which is obtained by the standard procedure of looking at the
{\em same}\/ collection of water particles (the ``system'')
at two nearby times, $t$ and $t+\Delta t$,
and writing that the rate of change of the system's total momentum
equals the total external force on the system.
The second equation states the hypothesized law of accretion for the raindrop:
here I shall consider the very general form
\be
   {dm \over dt} \;=\; \lambda m^\alpha v^\beta
 \label{eq.accretion}
\ee
where $\lambda>0$ is a constant
and $\alpha$ and $\beta$ are (almost) arbitrary exponents.
This form includes the two most commonly studied cases ---
namely the easy case
$(\alpha,\beta) = (\smfrac{2}{3},0)$
[accretion proportional to the surface area of a spherical raindrop,
 with resulting acceleration $g/4$]
and the hard case
$(\alpha,\beta) = (\smfrac{2}{3},1)$
[accretion proportional to the volume swept out,
 with resulting acceleration $g/7$] ---
but is much more general.\cite{note_other_cases}
I will show that all these problems can be solved by a unified technique.

First, a few preliminary remarks.
Since \reff{eq.newton}/\reff{eq.accretion} is a pair of
first-order differential equations for two unknown functions,
the general solution will contain two constants of integration.
Since this system is time-translation-invariant,
one of the constants of integration simply sets the origin of time.
The other constant of integration fixes the relation between the
initial mass and the initial velocity:
that is, it fixes the mass at the moment when the velocity has a specified
value (or vice versa).
The simplest solution arises by demanding that $m=0$ when $v=0$;
but we will make some partial progress toward finding
the general solution as well.

As mentioned above, all the cases with $\beta=0$ are easy:
the accretion equation~\reff{eq.accretion} can be solved immediately
for $m(t)$ [it is separable], and the Newtonian equation~\reff{eq.newton}
can then be solved for $v(t)$
[it is linear first-order with nonconstant coefficients].
The trouble arises when $\beta \neq 0$,
as now the equations \reff{eq.newton} and \reff{eq.accretion} are coupled.

To decouple them, we employ the technique mentioned earlier:
use the chain rule
\be
 {dv \over dt}  \;=\;  {dv \over dm} \: {dm \over dt}   \;,
 \label{eq.chain}
\ee
forget temporarily about time,
and instead consider the velocity $v$ to be a function of the mass $m$
(i.e.\ we temporarily use $m$ as the independent variable).
Inserting~\reff{eq.chain} into the Newtonian equation~\reff{eq.newton}
and using the accretion equation~\reff{eq.accretion}
to eliminate $dm/dt$ (which now multiplies both terms on the left-hand side),
we obtain
\be
    v^\beta {dv \over dm} \,+\, {v^{1+\beta} \over m}  \;=\;
   {g \over \lambda} \, m^{-\alpha}  \;.
 \label{eq.vm}
\ee
Making the change of variables
$w = v^{1+\beta}$,
we find\cite{note_betaminus1}
\be
   {dw \over dm} \,+\, {1+\beta \over m} w  \;=\;
   {(1+\beta)g \over \lambda} \, m^{-\alpha}  \;,
 \label{eq.wm}
\ee
which is a first-order linear differential equation
with nonconstant coefficients for the function $w(m)$.
The integrating factor is $m^{1+\beta}$,
and after standard manipulations we obtain
the general solution\cite{note_alpha=2+beta}
\be
   v  \;=\;
   \left[
   {(1+\beta)g \over (2+\beta-\alpha)\lambda} \, m^{1-\alpha}  \:+\:
   {C \over m^{1+\beta}}
   \right] ^{\! \scriptstyle 1 \over \scriptstyle 1+\beta}  \;.
\ee
It is most convenient to express the constant of integration $C$
in terms of the raindrop's mass $m_0$ at the moment its velocity is zero:
we have\cite{note_initial}
\be
   v  \;=\;
   K
   \, m^{\scriptstyle 1-\alpha \over \scriptstyle 1+\beta}
   \, \bigl[ 1 - (m_0/m)^{2+\beta-\alpha} \bigr]
                         ^{\scriptstyle 1 \over \scriptstyle 1+\beta}
 \label{eq.v_m_m0}
\ee
where $K = [(1+\beta)g / (2+\beta-\alpha)\lambda]^{1/(1+\beta)}$.

The simplest case is $m_0 = 0$.
Then we have\cite{note_m0=0}
\be
   m  \;=\; K' v^{\scriptstyle 1+\beta \over \scriptstyle1-\alpha}
 \label{eq.m_fn_v}
\ee
where $K'$ is a constant that we need not write out.
We now insert \reff{eq.m_fn_v}
into the Newtonian equation~\reff{eq.newton}.
Since each of the terms in this equation is linear in $m$,
the constant $K'$ drops out, and we get
\be
    {dv \over dt}  \;=\; {1-\alpha \over 2+\beta-\alpha} \, g  \;.
\ee
So we are done:  we have proved that the raindrop falls with
constant acceleration ${1-\alpha \over 2+\beta-\alpha} g$.
[When $(\alpha,\beta) = (\smfrac{2}{3},0)$ or $(\smfrac{2}{3},1)$,
 this gives $g/4$ or $g/7$, respectively.]
We can, if we wish, plug back in to get
$m(t) = K'' \, t^{(1+\beta)/(1-\alpha)}$.

When $m_0 \neq 0$, the best approach seems to be to insert~\reff{eq.v_m_m0}
into the accretion equation~\reff{eq.accretion} and integrate:
\be
   \int \! m^{- {\scriptstyle \alpha+\beta \over \scriptstyle 1+\beta}} \,
      \bigl[ 1 - (m_0/m)^{2+\beta-\alpha} \bigr]
           ^{- {\scriptstyle \beta \over \scriptstyle 1+\beta}}   \, dm
   \;=\;
   \lambda K^\beta \! \int \! dt
   \;.
 \label{eq.intm}
\ee
If $\beta = 0$, the integral is easy:  we get
\be
   m(t)  \;=\; m_0 \bigl[ 1 + (1-\alpha)\lambda t/m_0^{1-\alpha} \bigr]
                          ^{\scriptstyle 1 \over \scriptstyle 1-\alpha}
\ee
and
%
\be
   v(t)  \;=\; {1-\alpha \over 2-\alpha} gt
               \,+\,
               {m_0^{1-\alpha} g \over (2-\alpha)\lambda}
               \left\{ 1 - \bigl[ 1 + (1-\alpha)\lambda t/m_0^{1-\alpha} \bigr]
                             ^{-{\scriptstyle 1 \over \scriptstyle 1-\alpha}}
               \!\right\}
    \,.
\ee
If $\beta \neq 0$, the substitution $z = (m_0/m)^{2+\beta-\alpha}$
allows the left-hand side of~\reff{eq.intm}
to be expressed as an incomplete beta function:
\be
   -\, {1 \over 2+\beta-\alpha} \, 
     m_0^{\scriptstyle 1-\alpha \over \scriptstyle 1+\beta} \,
      B\bigl( (m_0/m)^{2+\beta-\alpha}; a,b \bigr)
\ee
with $a = -(1-\alpha)/[{(1+\beta)}(2+\beta-\alpha)]$ and $b=1/(1+\beta)$
[this can alternately be written as a hypergeometric function $\tfo$].
But it seems difficult to make further analytic progress.
One can in any case see from \reff{eq.intm} that the large-time behavior is
\be
   m(t)  \;=\; K'' \, t^{\scriptstyle 1+\beta \over \scriptstyle 1-\alpha}
               \left(1 + \sum_{k=1}^\infty a_k \,
                 t^{-k {\scriptstyle (1+\beta)(2+\beta-\alpha) \over
                        \scriptstyle 1-\alpha}}
               \right)
\ee
and hence
\be
   v(t)  \;=\; {1-\alpha \over 2+\beta-\alpha} \, g t
               \left(1 + \sum_{k=1}^\infty b_k \,
                 t^{-k {\scriptstyle (1+\beta)(2+\beta-\alpha) \over
                        \scriptstyle 1-\alpha}}
               \right)
  \,,
\ee
from which the coefficients $\{a_k\}$ and $\{b_k\}$
can be determined by substitution into \reff{eq.newton}/\reff{eq.accretion}.

\begin{acknowledgments}
This article is dedicated to the students in MATH 1302
at University College London who suffered through my problem sets.

My research is supported in part by
U.S.\ National Science Foundation grant PHY--0424082.
\end{acknowledgments}

\vfill


\begin{thebibliography}{99}
\bibitem{note_solve_Fxv}  For instance, if $F(x,v) = g(x) h(v)$,
then \reff{eq.dvdx} is separable
[in particular, when $F(x,v) = g(x)$
this gives the usual conservation-of-energy equation].

\quad
If $F(x,v) = g(x) v + h(x) v^2$,
then \reff{eq.dvdx} is a first-order linear equation
with nonconstant coefficients for the function $v(x)$;
and more generally, if $F(x,v) = g(x) v^{2-\alpha} + h(x) v^2$,
then \reff{eq.dvdx} is a first-order linear equation
for the function $v(x)^\alpha$.

\quad
Likewise, if $F(x,v) = v/[g(v) + h(v) x]$,
then \reff{eq.dvdx} can be turned upside-down to get
a first-order linear equation with nonconstant coefficients
for the inverse function $x(v)$;
and more generally, if $F(x,v) = v/[g(v) x^{1-\beta} + h(v) x]$,
then \reff{eq.dvdx} gives
a first-order linear equation
for the function $x(v)^\beta$.
\bibitem{note_other_cases}  At least two other cases of~\reff{eq.accretion}
correspond to physically realizable (albeit highly artificial) situations:
namely, $(\alpha,\beta) = (0,1)$ and $(1,1)$
arise when the raindrop is constrained (e.g.\ by a massless container)
to be a cylinder of fixed base and growing height
(resp.\ fixed height and growing base).
Note that the cylinder's base can here be of arbitrary shape;
it need not be circular.
\bibitem{note_betaminus1}  The case $\beta=-1$ needs to be treated separately,
yielding
$$ v \;=\; C m^{-1} \exp\!\left[ \displaystyle {g \over (1-\alpha)\lambda}
                               m^{1-\alpha}
                        \right]
   \,.
$$
Of course, all cases $\beta < 0$ are unphysical.
\bibitem{note_alpha=2+beta}  The case $\alpha = 2+\beta$
needs to be treated separately, yielding
$$ v \;=\; \left( {(1+\beta) g \over \lambda} \:
                  {\log(m/m_0) \over m^{1+\beta}}
           \right) ^{\! \scriptstyle 1 \over \scriptstyle 1+\beta}
   \:.
$$
Of course, all cases $\alpha > 1$ are probably unphysical.
\end{thebibliography}
\end{document}